\documentclass[%
 reprint,
 amsmath,amssymb,
 aps,
prmaterials,
floatfix,
longbibliography
]{revtex4-2}

\usepackage{appendix}
\usepackage{graphicx}
\usepackage{dcolumn}
\usepackage{bm}

\usepackage{amsmath}
\usepackage{verbatim}
\usepackage{siunitx}
\usepackage[colorlinks=true, allcolors=blue]{hyperref}
\usepackage[normalem]{ulem}

\begin{document}


\title{Generation of Ultrashort Ion Pulses from Ultrafast Electron-Stimulated Desorption}

\author{M.C. Chirita Mihaila}
  \affiliation{TU Wien, Institute of Applied Physics, 1040 Vienna, Austria}
  
\author{G.L. Szabo}
  \affiliation{TU Wien, Institute of Applied Physics, 1040 Vienna, Austria}

\author{A. Redl}
  \affiliation{TU Wien, Institute of Applied Physics, 1040 Vienna, Austria}
  
\author{M. Goldberger}
  \affiliation{TU Wien, Institute of Applied Physics, 1040 Vienna, Austria}

\author{A. Niggas}
  \affiliation{TU Wien, Institute of Applied Physics, 1040 Vienna, Austria}
  
\author{R.A. Wilhelm}
  \thanks{Corresponding author: wilhelm@iap.tuwien.ac.at}
  \affiliation{TU Wien, Institute of Applied Physics, 1040 Vienna, Austria}
  
\date{\today}

             
\begin{abstract}
We present an efficient method to produce laser-triggered proton pulses well below 500\,ps pulse width at keV energies. We use femtosecond photoelectron pulses emitted from a cathode to enable ultrafast electron-stimulated desorption of adsorbates on a stainless steel plate under ultrahigh vacuum conditions. While direct photoionization of atoms to form well-timed ion pulses can suffer from a laser-focus-limited large starting volume, in our method the 2D starting plane of the ions is defined with nanometer precision at a solid surface. We clearly outline how the method could be used in the future to efficiently produce ion beam pulses in the (sub-)picosecond range for pump-probe experiments with ions.
\end{abstract}

\maketitle

\section{\label{sec:intro}Introduction}

The interaction of energetic ions with solid surfaces is used in a plethora of applications ranging from nanolithography with focused beams~\cite{hoflich_roadmap_2023}, to polishing of surfaces in the production of x-ray mirrors~\cite{zhang_high-precision_2022} and secondary ion mass spectrometry in material analysis~\cite{fletcher_secondary_2013,wucher_laser_2013} among many others~\cite{wu_recent_2023,wilhelm_charge_2022}. The fundamental understanding of material modification and erosion by ions relies, however, mostly on simulations within the binary collision approximation (BCA)~\cite{hofsass_binary_2022,wilhelm_missing_2023,moller_shaping_2016} or molecular dynamics (MD) framework~\cite{schlueter_absence_2020,nordlund_primary_2018}. While the latter yields detailed insight into the time evolution of surface atom motion after ion impact~\cite{nordlund_historical_2019}, MD simulations are based on force fields between surface atoms typically tuned to material properties at room temperature~\cite{ostadhossein_reaxff_2017} or somewhat elevated temperatures~\cite{botu_machine_2017}. The collisional cascade triggered by the impacting ion and ultimately responsible for atom sputtering contains thousands of moving atoms at several 100\,eV to few keV of kinetic energy ($\sim 10^5-10^7$\,K). Furthermore, the Ziegler-Biersack-Littmark ion-target interaction potential, mostly used in typical MD simulations, is less accurate for ion-target systems with largely different masses~\cite{hofsass_simulation_2014} and, as a screened potential in the statistical atom picture, it describes in fact only neutral atom scattering neglecting the ion's charge state entirely~\cite{thomas_calculation_1927,fermi_statistische_1928}. While there are major improvements in MD using machine learning force fields~\cite{chmiela_towards_2018}, coupling to time-dependent density functional theory in the Ehrenfest dynamics approach~\cite{ojanpera_nonadiabatic_2012,ojanpera_electronic_2014,niggas_charge-exchange-dependent_2023}, improvements in GPU-based calculations allowing larger system sizes and longer simulation times~\cite{glaser_strong_2015}, ultimately full predictive power of these methods is not yet reached, clearly outlining that experimental benchmarks at the MD simulation timescale are needed.

In ion-solid experiments, the ultimate challenge is still to perform a truly time-resolved study to follow the surface dynamics after ion impact in real-time. Only then one would be able to falsify or verify models and simulations with which the wide realm of ion beam applications in materials science is described. What hinders these experimental studies so far is the lack of precise timing of ion pulses and the typical absence of time-synchronization to a second probing pulse in the powerful stroboscopic pump-probe scheme~\cite{zewail2000femtochemistry,krausz2009attosecond,zewail_four-dimensional_2010,feist2015quantum,wang2020coherent,tauchert2022polarized,tsarev2023nonlinear}. Substantial progress was made recently by introducing femtosecond high power laser systems for well-timed photoionization~\cite{golombek_generation_2021,kalkhoff_path_2023} resulting in 3-5\,ps proton pulses at MeV energies~\cite{dromey_picosecond_2016} or $\sim 18$\,ps Ne$^+$ pulses at keV energies~\cite{kalkhoff_path_2023,golombek_characterization_2020,breuers_concept_2019}. Still, MeV protons interact with a surface mostly by electronic excitations and not the application-relevant sputtering processes, and the 18\,ps Ne$^+$ pulses are produced with high optical laser powers exceeding $10^{15}$\,W/cm$^2$ in a miniaturized pulse buncher setup. 

Our approach discussed here comprises a $\sim 20$\,cm long beamline for versatile pump-probe experiments. We use ultrafast electrons generated at low optical power densities of $10^9-10^{10}$\,W/cm$^2$ to drive a Franck-Condon transition upon impact on a surface. This transition can shift the ground state of adsorbed particles from the surface into repulsive excited states. As such, the excited particles are pushed away from the surface and can desorb. Distinct constituents, including positively and negatively charged ions, as well as neutral particles are created. The method is known as electron-stimulated desorption (ESD)~\cite{menzel1964desorption,madey1970isotope,redhead1970ion,madey1971electron,menzel1975electron,drinkwine_electron-_1976,drinkwine1977electron,knotek1978ion,antoniewicz1980model} and was discussed when using ultrafast electrons in~\cite{backus_ultrafast_2007}. 

To make use of short ion pulses for time-resolved surface studies (ions as pump, laser as probe), three major criteria must be fulfilled: (1) Ion pulses must be significantly shorter than the time in which the process under investigation takes place. The process duration spans from picoseconds to $\sim10$\,ns~\cite{perez_picosecond_2024,choudhry_persistent_2023}. (2) The ion pulses must be well-synchronized to a pulsed laser for the pump-probe scheme with a laser-ion time jitter of less than the ion pulse duration. (3) The solid surface must allow many ion impacts before it degrades too much, still showing a large-enough response to the ion impact to be probed by the laser. Alternatively, the experimental approach needs to allow for a fine scanning of the ion pulses to expose a pristine part of a surface at each shot. Here we show a new, yet simple method for versatile production of ion pulses of many different ion species in line with (1) and readily compliant with (2). The method allows an ion transport from point of ionization to the pump-probe interaction spot over several centimeters making it possible to fit standard material samples and optical equipment for flexible measurement strategies properly addressing (3).

\section{\label{sec:setup}Experimental Setup}

\subsection{\label{sec:uebis}UEBIS}

\begin{figure}[h]
   \centering      
   \includegraphics[width=8.6cm]{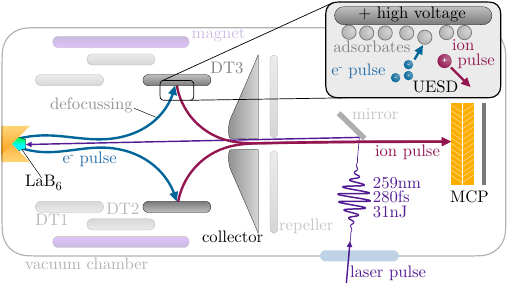} \caption{Sketch of the pulsed ion source, not to scale. A UV laser pulse enters the UEBIS and is focused onto a LaB$_6$ cathode to trigger ultrafast electron emission. The electrons are then guided towards DT3, which is made from of stainless steel. Here, the ions are desorbed forming pulses, accelerated and guided towards the MCP detection system. More detail on the experimental setup is given in Sec.~\ref{sec:setup}.}
   \label{fig1}
\end{figure}

We use a minitaturized (Ultrafast) Electron Beam Ion Source (UEBIS) from DIS Germany GmbH~\cite{d-i-s_germany_httpswwwdis-engproductsion-sources_nodate} where the standard thermal cathode is replaced by a LaB$_6$ photocathode (Kimball physics, ES-423E-9015, flat tip apex). The cathode emits pulsed electrons due to illumination with 290\,fs laser pulses of 259\,nm wavelength (see Fig.~\ref{fig1}). The UV laser pulses are produced from a 1035\,nm pulse through fourth harmonic generation utilizing two consecutive $\beta$-barium oxide (BBO) non-linear crystals. The UV laser beam is guided through a half-wave plate (HWP) and polarizer, followed by a second HWP to freely adjust the power and polarization. Finally, the UV laser beam is focused (FWHM $\sim 50$\,$\mu$m) by a lens with focal length $f= 50$\,cm on the cathode of the EBIS. For all experiments with pulsed ions, the UV laser repetition rate is set to 100\,kHz and the laser pulse energy $E_L \sim 31$\,nJ. 

The photoelectrons from the LaB$_6$ cathode are focussed by an electric field between cathode at the first drift tube (DT1) of 3.5\,kV/mm and the axial magnetic field of 100\,mT of the UEBIS. The DT ensemble is surrounded by Ni-coated SmCo permanent magnets inside the vacuum chamber. Ions produced in the ion source are only weakly affected by the $B$-field due to their comparatively high mass. After passing the DT region the electrons are dumped at a collector plate with the help of a negative repeller voltage (cf. Fig.~\ref{fig1}). Switching the repeller voltage off allows the extraction of (at least a fraction of) the electrons in each pulse and we measured $\sim 10^3$\,electrons per laser pulse at the microchannel plate (MCP). The electron pulse population \textit{inside} the UEBIS, where ionization occurs, might be higher than the value we determine outside the UEBIS on the MCP detector. Even in standard operation of an EBIS a fraction of typically $\sim 1/1000$ of the electrons impact the drift tubes in front of the collector contributing to an (typically) unwanted blind current. We make use of exactly these stray electrons to drive (ultrafast) electron-stimulated desorption (UESD) at the surface of the third drift tube (DT3). Note that the impact energy of the electrons at DT3 is $eU_{DT3}+e|U_{cat}|$, where $U_{DT3}$ is the positive potential of DT3 and $U_{cat}=-0.8$\,kV is the potential of the cathode. The ions are accelerated in between DT3 and the grounded collector and by the (negative) front potential of $-2.1$\,kV of the MCP detector. A description of the UESD process is given in the next section. Typical high voltage conditions are: $U_{cat}=-350$\,V, $U_{DT1}=+3300$\,V, $U_{DT2}=+5000$\,V, $U_{DT3}=+8500$\,V, $U_{repeller}=-200$\,V, $U_{MCP,front}=-2100$\,V.

Positioning of the UV laser beam at the cathode is done by first measuring the black-body radiation intensity from Ohmic heating the LaB$_6$ crystal after a collimator and guiding the laser beam on the same path with the help of two irises, as described in~\cite{Kozak2018b}. The surface of the mirror in the vacuum chamber is gold coated to avoid charging effects influencing the passing ion pulse.

Desorbed ions from DT3 are immediately accelerated by applying a positive potential to the DT3, collimated by the electron collector and repeller, and are time-stamped using the Microchannel Plate (MCP) assembly in coincidence with a fixed trigger signal from a photo diode (PD, Femto, FWPR-20-IN-FST) picking up part of the initial IR laser intensity.

\subsection{Data Acquisition} 

The data acquisition is initiated with a fraction of the Infrared (IR) laser intensity being directed towards the PD (trigger signal), which results in an analog output signal. This signal is then fed into a pico-timing discriminator (Ortec 9307). Subsequently, this digital signal serves as the start signal for the time-to-amplitude converter (TAC, Ortec 567). 

In the second part of the process, the analog signal coming from the MCP traverses through an AC signal decoupler (Roentdek, HFSD-SMA SHV 10\,k$\Omega$), followed by a fast 2\,GHz amplifier (Femto, HSA-Y-2-40). Due to the variable amplitude, the signal is analyzed with a constant-fraction discriminator (Roentdek, CFD1x). The CFD output serves as the TAC stop signal.

The last stage involves analyzing the output signal from the TAC using a multi-channel analyzer (MCA, CAEN N957) to generate a TOF histogram. The MCA utilizes 13\,bit on $0-10$\,V input signals. At a TAC range of 50\,ns ($\hat{=} 10$\,V) this corresponds to $\sim 6$\,ps MCA binning, well below the system jitter of 330\,ps. At 1000\,ns TAC time base (see Fig.~\ref{fig4}), however, the MCA binning is 120\,ps which does allow peak width determination only with limited accuracy.

To corroborate our experimental results and to pinpoint the point of ionization for the ions we observe, we performed charge particle trajectory simulations with the SIMION code~\cite{simion_code_package_httpssimioncom_nodate}. The SIMION simulations confirm the electron pulse trajectories impacting the DT3 and the timing structure of pulsed protons starting from the DT3 surface (see Appendix Figs.~\ref{fig5} and \ref{fig7}).

The ion source is operated at a base pressure of $\sim 5\times 10^{-9}$\,mbar without baking the chamber and no additional gas is fed into the vacuum vessel.

\section{\label{sec:results}Results}

\begin{figure}[h]
   \centering      
   \includegraphics{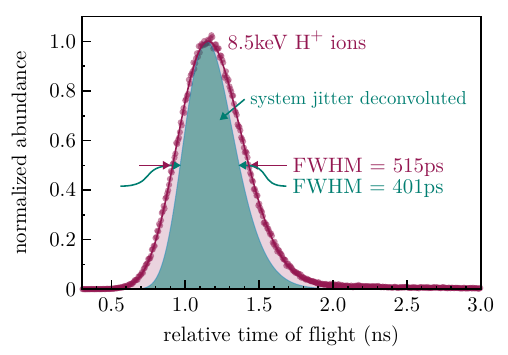} 
   \caption{The arrival time difference between the pulsed ions at the MCP and a fixed trigger signal from the PD yields the histogram. The zero reference time is chosen arbitrarily. The histogram is fit with a LogNormal distribution revealing a FWHM of 515\,ps of the raw data and 401\,ps after deconvolution with the determined system jitter. The voltage applied to DT3 is set to $+8.5$\,kV and the MCP front voltage to $-2.1$\,kV.}
   \label{fig2}
\end{figure}

The shortest proton pulse time distribution at 8.5\,keV energy is shown in Fig.~\ref{fig2} and follows a LogNormal distribution reasonably well as expected from a skewed Gaussian pulse width where the shortest times (lower bound) are pinned by the time of flight (TOF) of shortest possible trajectory (shortest possible travel time considering the electric field geometry). It should be noted that the integral of this timing peak is consistent with 0.015\,ions per laser pulse ($10^3$\,ions/s for 100\,kHz laser repetition rate), i.e., the pulse width shown here is the arrival time distribution of single ions and reflects the ion timing uncertainty in a future pump-probe experiment. The ion output per laser pulse increases linear with laser power reaching about 0.035\,ions/laser pulse at $E_L \sim 67$\,nJ. The raw TOF spectrum (red), however, is influenced by a timing jitter from the MCP itself, the constant-fraction-discriminator (CFD), the time-to-amplitude converter (TAC), and the photo-diode, i.e., all 'active' electronic components used for TOF measurement. Slightly defocussing the UV light in the vacuum chamber allows to trigger the MCP by reflected UV photons having a timing distribution at the MCP only given by different reflection points in the source (max. 5\,mm line-of-sight length difference yields max. 17\,ps timing width). This laser timing profile is measured and shows a Gaussian distribution with a FWHM of 330\,ps. Consequently the raw TOF spectrum of the protons is deconvoluted with a Gaussian kernel of FWHM of 330\,ps resulting in the green profile in Fig.~\ref{fig2} with a FWHM of 401\,ps.

\begin{figure}[h]
   \centering      
   \includegraphics{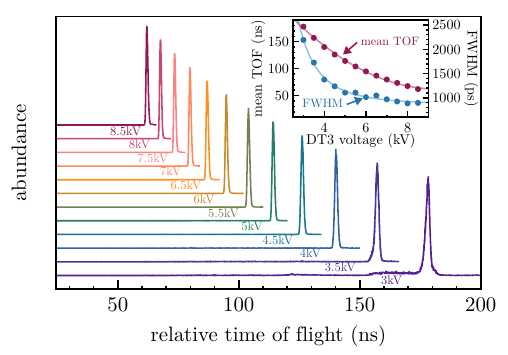}
   \caption{Time of flight spectra of the H$^+$ peak for different voltages applied to DT3 indicated in the graph. The same arbitrarily chosen zero reference time is substracted from each spectrum. The mean TOFs and peak widths are shown in the inset. From the mean TOF shift we identify the ion species to be protons and the peak width decreases with a power law reaching a minimum of 852\,ps for the ion source conditions used here, which are different from the ones of Fig.~\ref{fig2} (see text). No peak deconvolution was done here.}
   \label{fig3}
\end{figure}

In order to clearly assign the TOF peak in Fig.~\ref{fig2} to H$^+$ starting at the DT3 surface, we swept the voltage applied to DT3 and recorded the corresponding peak shift shown in Fig.~\ref{fig3}. From the shift of the distribution mean value with voltage applied to DT3 we can extract the ion mass to be 1\,u. Note that the DT1 and DT2 voltages for the series in Fig.~\ref{fig3} are intentionally different than the ones used for Fig.~\ref{fig2} showing that the timing performance of the UEBIS depends significantly on the particular electron trajectories and impact points being the start triggers for ionization in the UESD scheme.

\begin{figure}[h]
   \centering      
   \includegraphics{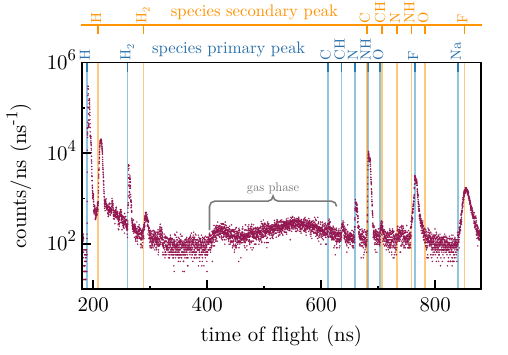} \caption{Time of flight spectrum of different ion species desorbed by ultrafast electrons. While the H$^+$ peak has the highest amplitude and smallest width, several different ion species ranging from H$_2^+$ to heavier molecular species like HN$^+$ are also desorbed from the DT3 surface. Moreover, a broader peak around $500 \pm 100$\,ns arises from the ionization of the residual hydrogen gas within the three drift tubes. Primary and secondary peaks for ion species obtained by a SIMION simulation are indicated with vertical lines which fit the onset of each TOF peak.}
   \label{fig4}
\end{figure}

Fig.~\ref{fig4} shows a TOF spectrum with larger time base, which allows to obtain more peaks at longer TOF. Most importantly, the UESD scheme allows also the formation of heavier ions like N$^+$ and molecular ions. Between 400 and 600\,ns, a broad feature can be seen, which we traced to ions formed by electron impact ionization of the rest gas within the UEBIS drift tubes, because the overall TOF is consistent with the expected kinetic energies and flight distances for monomer and molecular hydrogen ions. Between 180 and 300\,ns, there are four peaks which all shift with voltage applied to DT3 (cf. Fig.~\ref{fig3}), where we identify the first two as H$^+$ and the latter two as H$_2^+$. Note, that also the other peaks for higher mass shift with voltage at DT3. To clarify the double peak structure per mass we consulted SIMION simulations, which reveal that only two distinct axial positions on the DT3 surface can lead to extraction of ions from the UEBIS. Other points of UESD along DT3 lead to trajectories which cannot pass the collector and repeller, which effectively form a collimator (see Appendix Fig.~\ref{fig7}).

\section{\label{sec:discussion}Discussion}

\subsection{\label{sec:esd}Electron-Induced Desorption Ionisation}

The rate at which particles desorb from the surface scales linearly with the intensity of impinging electrons and the abundance of the specific adsorbed entities present on the surface~\cite{menzel1975electron}. In our case, thermal desorption can be excluded due to the low (cw-equivalent) pulsed electron current.  

The quantum mechanical description of ESD has been investigated in~\cite{brenig1976quantum,schuck1982quantum}. Briefly, we follow~\cite{menzel1964desorption,redhead1964interaction,nishijima1970electron,madey1971electron}, and consider the classical probability for ion desorption $P_i(x_0)$ of adsorbed neutrals on a surface upon electron impact is given by
\begin{equation}
    P_i(x_0) = \mathrm{e}^{-\int_{x_0}^{\infty} \frac{N_r(x)}{v(x)}dx} ,
    \label{Probability for ionic ESD} 
\end{equation}
where $x_0$ is the distance from the surface, $N_r(x)$ is the Auger neutralization rate in units of $s^{-1}$, $v(x) = \sqrt{2[V(x_0)-V(x)]/m_i} $ is the velocity at which the ion travels from the surface post-excitation, $V(x)$ is Born-Mayer repulsive potential and $m_i$ is the ion mass. Considering an exponential decay of probability for the Auger neutralization with distance $N_r(x) \propto e^{-ax}$ and $V(x) \propto e^{-bx}$ with some characteristic lengths $a$ and $b$, we see that the probability to find a charged desorbed atom or molecule is larger for light elements travelling with higher velocity and therefore reducing the likelihood for Auger neutralization. Further, not seen from Eq.~(\ref{Probability for ionic ESD}), the neutralization rate is smaller for non-resonant neutralization than for resonant one~\cite{nedeljkovic_final_2009,hagstrum_ion-neutralization_1966}, i.e., ions with high ionization potential are less likely to non-resonantly neutralize with an electron from the surface Fermi edge. With both arguments, we can understand that H$^+$ is the most abundant ionic species in Fig.~\ref{fig4}. Note, that also the sticking of adsorbates to a substrate material and the corresponding adsorption energies are element specific.
Another important aspect for short pulse formation is the initial velocity spread leading to dispersion over a longer ion trajectory towards a sample surface. In experiments~\cite{redhead1964interaction,nishijima1970electron,madey1971electron}, it has been observed that O$^+$ desorbs with an energy spread of $\Delta E \sim 3-5$\,eV. An O$^+$ pulse traveling at a speed of $v = 3.2 \times 10^5$\,m/s (8.5\,keV) with $\Delta v= 94$\,m/s ($\Delta E = 5$\,eV) disperses at a rate of $\sim 9.2$\,ps/cm. Interestingly, desorbed H$^+$ ions from tungsten covered with H$_2$ adsorbates have $\Delta E \sim 1$\,eV, while H$^+$ from adsorbed H$_2$O have $\Delta E \sim 2$\,eV~\cite{nishijima1970electron}. For H$^+$ pulses with initial energy of 8.5\,keV $\pm$ 1\,eV we estimate a dispersion of $\sim 19$\,ps in our setup over 20.5\,cm of flight distance. To minimize geometrical spreading of ion trajectories leading to different flight paths as well as TOFs, it is beneficial to apply a large electric field at the point of ionization. Then, desorbed ions are pulled along the electric field lines and accelerated quickly further reducing dispersion, as seen in Fig.~\ref{fig3}.

\subsection{Off-axis Electron Trajectories}

To follow the electron trajectories towards the DT3 surface, i.e., their spreading in radial direction, we need to consider that (a) the laser focus ($\sim 50$\,$\mu$m) is larger than the LaB$_6$ flat top size ($\sim 15$\,$\mu$m) and hence electrons are also emitted off-axis from the conical side surfaces of the cathode. The cathode is positioned such that the flat top surface is at the Brillouin point of the axial magnetic field ($\sim 100$\,mT). For electron emission from the conical side surface, the magnetic field is $\neq 0$ which then leads to a oscillatory motion of the electrons around the central axis. At DT3, also due to the high voltage applied, these off-axis electrons hit the surface and lead to UESD. In Appendix~\ref{sec:electrons} and \ref{sec:simion_ion} we present in detail charged-particle trajectory simulations confirming this behavior together with laser-polarization-dependent measurements which indicate that the UESD process is triggered from electrons originating from a surface with an appreciable inclination angle with respect to the incoming laser beam.
The starting area for the electrons from the side surfaces of the cathode is determined by an annulus with an inner diameter of $\sim 15$\,$\mu$m for the given LaB$_6$ flat top size and an outer diameter of $\sim 50$\,$\mu$m given by the laser focus and amounts therefore to $\sim 1800$\,$\mu$m$^2$. An electron pulse of $10^5$\,electrons would then yield an available starting area of $10^4$\,nm$^2$ per electron. Consequently, for our given geometry, dispersion due to Coulombic repulsion between the electrons for any given pulse is negligible.

 
\section{\label{sec:conclusion}Conclusion and Outlook}

In our experiment, the setup is designed with a long flight distance ($\sim 20$\,cm) to make the alignment easier. This distance can be reduced in future designs to minimize the pulse broadening of ions. Following the estimation from the initial ESD energy spread for O$^+$ above, a reduction by $\sim 10$\,cm would reduce the dispersion by $\sim 90$\,ps, i.e., especially for heavier species this needs to be considered. For hydrogen the estimated dispersion by the initial kinetic energy spread amounts in total to only 19\,ps in our setup, thus the system size is not the limiting factor for shorter pulses. The initial position spread of the desorbed H$^+$, however, could be narrowed down by decreasing the diameter of the electron focal spot on the metallic plate or further narrowing the collimator formed by the collector and repeller electrodes. Here, UESD shows a promising pathway towards integration into ultrafast scanning electron microscopes, where a few nanometer focus of ultrafast pulsed electrons is readily available and would then allow a nanometer precise ionization point in 3D space from which ion pulses with ultimate timing performance could be formed. 

It should be noted that we detect the ions on a 2-inch MCP detector without spatial resolution. Different ionization points at DT3 lead to different trajectories through the collector (cf. Appendix Fig.~\ref{fig7}(b)) and finally different impact points on the MCP. Using a position-resolved MCP detector would allow to disentangle the timing and position (i.e., trajectory) dependence and we expect significantly shorter pulses by further geometric filtering.

More advanced pulse compression strategies to propel our technique into the single-digit picosecond regime for protons and into the sub-ns regime for heavy species could be considered as well. Among these methods, utilizing radio frequency (RF) fields~\cite{van2010compression} could offer promising prospects. Even in non-space charge limited regimes, RF compression can effectively reduce pulse duration~\cite{gliserin_compression_2012} at the cost of introducing a broad spectrum of kinetic energies within the compressed pulse. This broadening occurs because the RF field imparts different energy changes to particles at different positions within the pulse, leading to a spread in kinetic energy. Similarly, the integration of photon-induced near-field electron microscopy (PINEM) technique~\cite{barwick2009photon,lundeberg2017tuning,akbari2022optical,morimoto2023attosecond}, known for its efficiency in pulse manipulation of electrons, could potentially be adapted also to reduce the pulse length of ions effectively.

Importantly, the UESD process is versatile in formation of different ion species. It is noteworthy that UESD can produce positive ion pulses, as demonstrated in our study, but also negative ion and neutral atom pulses~\cite{szymonski_electron_1992,hock_electron_1978}, which remain primarily unexplored, presenting a rich field for further investigation. The metallic electrode at which UESD takes place could be cooled by LN$_2$ or placed in a cryostat in order to reduce the initial temperature and velocity spread of desorbed species as well as to serve as a reservoir of atomic and molecular species adsorbing at low temperatures. Using atomically flat (single crystalline) surfaces well prepared under UHV conditions and loaded with a gaseous species adsorbing at well defined atomic sites, would allow to calculate the initial velocity for ESD and minimizing the spread by preferentially allowing only one specific type of adsorption site and atomic orientation of the adsorbate at the surface.

For a future pump-probe experiment UESD is a promising process while achieving picosecond laser synchronization at a sample for the ion TOF of 0.1-1\,$\mu$s is still challenging. Technical solutions such as optical cavities using parabolic mirrors can enable multiple laser turns in a $\sim 30$\,cm cavity~\cite{herriott_folded_1965}. By combining these with a commercial picosecond delay stage, delays up to 200\,ns with less than 100\,ps jitter are achievable. For even longer delays one can utilize subsequent laser pulses at a 5\,MHz laser repetition rate to add a constant 200\,ns delay.

In this study, we have demonstrated the generation of picosecond ion pulses ($10^3$ H$^+$ pulsed (400\,ps) ions per second at 100\,kHz laser repetitian rate) utilizing ultrashort electron pulses. The ion species produced in the current setup are determined by the residual gas adsorbed on the DT3 of the UEBIS. This signifies a successful initial foray into generating ultrafast ion pulses with a well-defined starting position, setting the stage for further development and refinement of this approach.
In conclusion, the UESD methodology presented in this study signifies a promising evolution in the generation of short-pulsed ion sources. It holds substantial potential for application in future pump-probe measurements and a technological platform to be readily implemented in existing ultrafast electron microscopes, thus providing a complete dynamical picture of molecular and atomic interactions in real time.

\begin{acknowledgments}
We thank A.S. Grossek and J. Fries for support with the SIMION calculations. This research was funded in part by the Austrian Science Fund (FWF) (Grant DOI 10.55776/Y1174). For open access purposes, the author has applied a CC BY public copyright license to any author accepted manuscript version arising from this submission. The authors declare no conflicts of interest. 
\end{acknowledgments}

\newpage
\appendix

\section{\label{sec:electrons} Pulsed Electron Signal}

To operate the UEBIS in pulsed electron detection mode, the front plate of the MCP is grounded, while the back plate is set to a positive voltage of 2.1\,kV and for measuring the DC-equivalent output (charge current measurement) the backside does not need to be biased. In this configuration, the specific voltages applied to the electrodes of the UEBIS are set for maximal electron current measured on the MCP front side.

The DC-equivalent current generated by photo-electrons from the LaB$_6$ cathode has been measured by connecting a picoamperemeter at the front side of the MCP. We detect a linear increase of electron output from the ion source with increasing $E_L$, reaching about 2,200\,e$^-$/pulse for $E_L = 15$\,nJ. Note that this is only a fraction of the electrons populating a pulse inside the source volume (see discussion above).  

\begin{figure}[h]
   \centering      
   \includegraphics{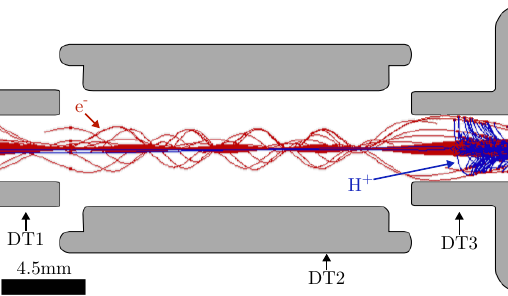} \caption{Simion simulated electron trajectories inside the UEBIS. Electrons starting off-axis (not shown) oscillate around the optical axis and are predominantly defocussed at DT1 and DT3 where their impact can trigger UESD.}
   \label{fig5}
\end{figure}

We performed simulations of both the electron and ion trajectories inside the ion source with the commercial code SIMION~\cite{simion_code_package_httpssimioncom_nodate}, including a fully detailed geometry of the ion source as well as the correct magnetic field configuration. Figure~\ref{fig5} shows a cut of the DT section with about 1000 electron trajectories. One can see that, on average, the electron trajectories get radially compressed by the magnetic field. However, electrons originating at larger radii than the LaB$_6$ flat top, i.e., from the conical side walls, oscillate around the beam axis. 

\begin{figure}[h]
   \centering      
   \includegraphics{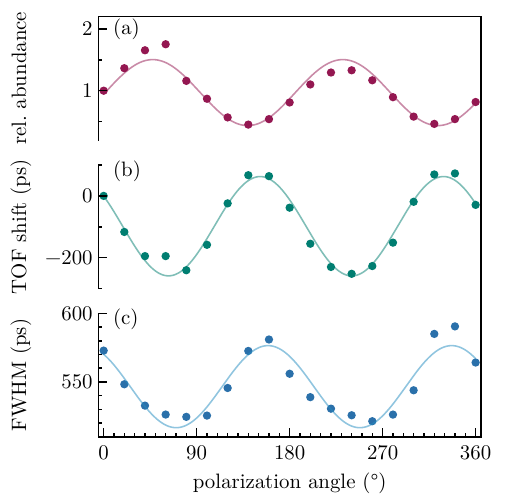}
   \caption{Polarisation-dependent H$^+$ signal: (a) The relative amplitude of the H$^+$ signal is recorded as a function of HWP rotation angle. (b) The mean value of the H$^+$ TOF peak shifts with laser polarization. (c) The measured FWHM as a function of the HWP rotation angle.}
   \label{fig6}
\end{figure}

Figure~\ref{fig6} shows that the timing peak amplitude and FWHM of the measured H$^+$ ion pulses can be enhanced by optimizing the laser beam polarisation with a HWP. It is well known~\cite{VENUS1983452,caruso2017development} that the pulsed electron emission is maximized when the laser beam polarisation is parallel to the photocathode, which finally results in more H$^+$ ions being desorbed from DT3. While under normal laser incidence on the cathode flat top, no polarization dependence is expected, it indeed is for the conical side walls. Thus, the larger laser focus leads to off-axis emission of electrons, which are also sensitive to the laser polarization direction.

\section{\label{sec:simion_ion} Charged Particle Trajectory Simulations for H$^+$}

\begin{figure}[h]
   \centering      
   \includegraphics{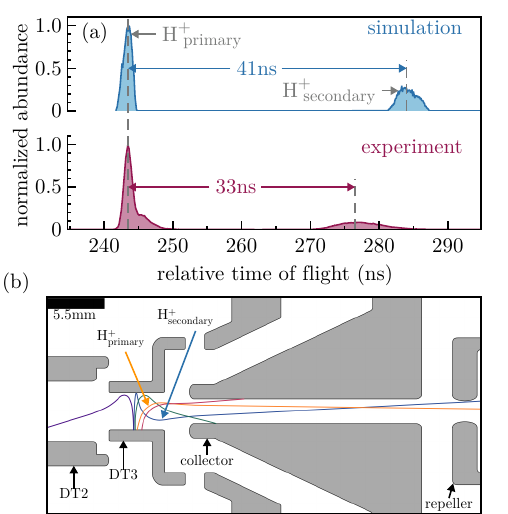}
   \caption{(a) shows the TOF distribution for the primary and secondary H$^+$ peaks observed in the simulations (top) with a time separation of 41\,ns and the experiment (bottom) with a time separation of 33\,ns, but very similar widths. (b) shows a snapshot of the SIMION simulations which yield the results in (a) presenting exemplary ion trajectories which get extracted from the source or filtered by the collector electrode.}
   \label{fig7}
\end{figure}

We simulated H$^+$ trajectories for ions starting at the inner DT3 surface along one line employing the axial symmetry of the system. The ions are emitted with a uniform energy distribution ranging from $0 - 1$\,eV. Initial emission is assumed to be normal to the surface for simplification.
Fig~\ref{fig7}(b) shows four individual ion trajectories as an example. Starting positions on the left side of DT3 (violet) lead to trajectories towards the cathode due to the lower positive voltage of DT2. The following starting positions to the right lead to extraction towards the exit of the UEBIS. However, the ions first move in the radial direction and only then turn around to exit the ion source. If the turning points of the trajectories are inside the collector and radially farther out than the collector's inner diameter, the ions are stopped. Thus, the collector is essentially a geometrical filter for specific starting positions. Under the conditions we observe the smallest TOF distribution in the experiment we also find two H$^+$ peaks both following the voltage applied to DT3 as shown, for example, in Fig.~\ref{fig4}. These two peaks can also be found in the SIMION simulations (see Fig.~\ref{fig7}(a)), because two distinct starting positions lead to extraction of ions, where they are separated in TOF by 41\,ns in the simulation and 33\,ns in the experiment with virtually the same peak widths of the first peak. We want to emphasize that SIMION allows an identification of the main reason for the double peak structure, but we abstain from further tuning the simulation for a better match with experiment due to missing information on the exact starting angle and energy distribution.

\section{\label{sec:jitter} System Jitter Determination}

With the UV pulse reflected in our vacuum chamber, we can also trigger the MCP, and the corresponding timing signal (MCP timing with respect to PD timing signal) should ideally be a $\delta$-function. Reducing the UV intensity such that the MCP output from UV reflections becomes comparable to the signal amplitude we observe for the H$^+$ (the CFD internal jitter depends on the input signal amplitude) allows a direct determination of the time broadening from the MCP and electronics we should expect for H$^+$. Using the UV light reflection, we determine an electronic system jitter of $\sim 330$\,ps from the Gaussian peak width (FWHM) measured between the MCP and PD signals.


\section{\label{sec:tof_calib} TOF to mass calibration}

The ions formed by UESD at DT3 are constantly subject to accelerating electric fields. Under the conditions of Fig.~\ref{fig4}, the ions are accelerated in between DT3 (on +8.5\,kV) and the collector (on 0\,V), in between the collector and repeller (on -200\,V) and towards the MCP (front on -2.1\,kV). To calibrate the TOF to mass, we consulted SIMION and extracted calibration curves following a squareroot dependence for both the primary and secondary peak for each mass. Still, to fit the mass spectrum in Fig.~\ref{fig4} a constant timing offset of 17.5\,ns was added for all masses, which is a result of delays in the electronic components used in the experiment. Furthermore, the TOF extracted from SIMION for the second peak needed to be reduced by 6.3\% for each mass to fit the experimental spectrum. Since the secondary peaks for H and H$_2$ were identified independently through a sweep of the DT3 voltage (cf. Fig.~\ref{fig3}), and by the fact that minor discrepancies between experiment and SIMION might still be expected, we are confident that this mass calibration holds also for the secondary peak. Note that the SIMION TOFs should be at the onset of each experimental TOF peak since SIMION considers ideal conditions and any imperfection in experiment will lead to longer, but never shorter TOFs.

\bibliography{references/literatur,references/references_wilhelm}

\end{document}